\documentclass[showpacs,amsfonts]{revtex4}
\usepackage{amsmath}
\usepackage{latexsym}
\usepackage{float}
\usepackage{amssymb}
\usepackage{graphicx}
\usepackage{textcomp}
\usepackage{hyperref}
\usepackage{graphicx}
\begin{document}

\def\be{\begin{equation}}
\def\ee{\end{equation}}     
\def\bfi{\begin{figure}}
\def\efi{\end{figure}}
\def\bea{\begin{eqnarray}}
\def\eea{\end{eqnarray}}

\newcommand{\bra}{\langle}
\newcommand{\ket}{\rangle}
\newcommand{\vk}{\vec{k}}
\newcommand{\rphi}{\widehat{\varphi}}
\newcommand{\iphi}{\widetilde{\varphi}}
\newcommand{\reta}{\widehat{\eta}}
\newcommand{\ieta}{\widetilde{\eta}}
\newcommand{\vphi}{\boldsymbol{\varphi}}
\newcommand{\eRk}{\emph{R}_{\vec{k},\eta}}
\newcommand{\eIk}{\emph{I}_{\vec{k},\eta}}

\title{Singular behavior of fluctuations in a relaxation process}

\author{Federico Corberi$^{1}$, Giuseppe Gonnella$^{2}$,
and Antonio Piscitelli$^{3}$} 
\affiliation {$^{1}$Dipartimento di Fisica ``E.R. Caianiello'', 
and CNISM, Unit\`a di Salerno,
Universit\`a  di Salerno, 
via Ponte don Melillo, 84084 Fisciano (SA), Italy. \\
$^{2}$Dipartimento di Fisica, Universit\`a di Bari and
INFN, Sezione di Bari, via Amendola 173, 70126 Bari, Italy. \\
$^{3}$Division of Physical Sciences, School of Physical and 
Mathematical Sciences, Nanyang Technological University, 
21 Nanyang Link, 637371, Singapore.}

\begin{abstract}

Carrying out explicitly the computation in a paradigmatic model of
non-interacting systems, the Gaussian Model, we 
show the existence of a singular point in the 
probability distribution $P(M)$ of an extensive variable $M$. 
Interpreting $P(M)$ as a thermodynamic potential 
of a dual system obtained from the original one by applying a constraint, 
we discuss how the non-analytical point of $P(M)$ is the counterpart 
of a phase-transition in the companion system. 
We show the generality of such mechanism by considering both 
the system in equilibrium or in the
non-equilibrium state following a temperature quench.
  

\end{abstract}

\maketitle
e-mail address: corberi@sa.infn.it

\section{Introduction}
\label{I}

Phase transitions are described by thermodynamic functions displaying singular behavior. 
For many complex systems,  like  inhomogeneous or disordered systems,
the basic mechanisms of phase transitions are still under debate.
On the other hand, there is  a large class of systems where   
the occurrence of a phase-transition can be ascribed to the existence
of a constraint, acting as an effective interaction for otherwise independent variables.
For instance, it is well known that
in bosons with an unrestricted number, as photons or
phonons, phase transitions do not show up. But the situation changes
radically when particles with a conserved number are considered,
leading to the remarkable phenomenon of the Bose-Einstein 
condensation (BEC)~\cite{Huang}. 
The condensation  mechanism, with  a certain sector of the phase space (the zero wavevector model in BEC) 
 becoming macroscopically populated, 
 is typical of  phase transition in constrained systems. Another worth example
is provided  by the Spherical model of Berlin and Kac~\cite{BK}, 
 obtained by constraining the order parameter field of the Gaussian model on a hypersphere of radius $S$~\cite{BK,LN}.
While the Gaussian model  is paradigmatic of a non interacting system~\cite{Goldenfeld} with well-known 
trivial equilibrium properties, the Spherical model shows  a highly
non-trivial behavior, characterized by a second-order phase transition in $d=3$. 

Constraints also appear in a conceptually different context,
when one wants to evaluate the  probability
of observing an extremely unlikely 
value $M$ of a macrovariable ${\cal M}$, as
for instance the energy in a canonical setting, due to a 
rare fluctuation of a thermodynamic system. 
Loosely speaking, in a sense that finds a clear explanation in the context of the {\it large deviation theory} \cite{Touchette} and that will better qualified in section~\ref{II}, 
the measurement of such probability can be 
regarded as a constraint applied on the system,
since this basically amounts to keep the configurations 
where the constraint ${\cal M}=M$ is fulfilled, 
discarding the others. 
Then, recalling the previous discussion, even if the average properties of the 
system under study are trivial, it can appear not surprising that the measurement of the probability distribution
of some of its macrovariables may show singular points.
However, only very recently, the occurrence of singularities in the large deviation functions 
of a number of different models has been recognized~\cite{schutz,Kafri,FL,US,Gambassi,Evans}
and interpreted in terms of a condensation mechanism. 

Recently~\cite{noi} we have studied the occurrence of a non-analytical
behavior in the 
probability distribution
of macrovariables in the context of the Gaussian model.
Choosing ${\cal M}={\cal S}$, the order parameter variance, this amounts
to impose the Berlin-Kac constraint, as said above.
The purpose of this paper is, after reviewing some of the results of~\cite{noi},
to discuss the singular behavior of 
the probability distribution of ${\cal S}$ not only in equilibrium
but also, by 
considering the relaxation following a temperature quench,
in the largely unknown area of the non-equilibrium processes
without time translation invariance~\cite{ritort,Ciliberto}. 
 
The paper is organized as follows: In section~\ref{II} we describe on general
grounds the relation between the probability of fluctuations of macrovariables
and the application of a constraint and, in 
section~\ref{III}, how such probabilities can be 
computed by saddle point techniques in the
large-volume limit.
The Gaussian model is introduced in section~\ref{IV}. This is the central 
section of the paper, where the probability distribution of
a particular macrovariable is explicitly determined (Secs.~\ref{V}, \ref{stde})
and its non-analytical behavior is discussed (Sec.~\ref{sing}). 
This leads to the determination of a phase-diagram,
namely the parameter region
where condensation of fluctuations occurs, in section~\ref{VI}.
Finally, in section~\ref{VII} we conclude by a discussion of some open points 
and the perspectives of future research on the subject.

\section{Probability distributions of macrovariables and constraints}
\label{II}

Let us consider a thermodynamic system whose microscopic degrees of freedom
we denote by $\varphi = \{\varphi_i\}$, where $i$ labels each of such variables.
For instance 
$\varphi _i \equiv \vec S_i$ could be the spin on the sites $i$ of a lattice 
in the case of a magnetic system.

Let $P(\varphi,J)$ be the probability distribution
of the microstates in the presence of certain control parameters $J$,
such as volume and temperature and, if the system is not in equilibrium,
time.
A generic random variable, like the energy of the system in contact 
with a bath, is a function ${\cal M}(\varphi)$ of the representative point
in the phase-space $\Omega$.  
The probability to observe a certain value $M$ of such fluctuating quantity
can be formally written as
\be 
P(M,J) = \int_{\Omega} d\varphi \, P(\varphi,J)\delta (M-{\cal M}(\varphi)).
\label{gen.3}
\ee

In the case of equilibrium states, the expression on the r.h.s.
of this equation can be readily 
interpreted as the partition function of a new system,
whose microstates $\varphi$ occur with probability  
\be
P(\varphi,M,J) = \frac{1}{P(M,J)} P(\varphi,J)\delta (M-{\cal M}(\varphi)).
\label{gen.2}
\ee
This system is 
obtained from the original one by fixing 
\be 
{\cal M}(\varphi)=M.
\label{constraint}
\ee
This constraint could be, in specific examples,  the conservation 
of the number of particles
in a bosonic gas or the restriction on the hypersphere of the order-parameter
field in the Spherical model.

Moving the value of $M$ it may happen that a critical point $M_c$ is 
crossed in the constrained model. Resorting again to the previous examples,
by fixing all the other control parameters (among which temperature),
BEC is observed upon raising the free bosons number above a certain
value, or the ferromagnetic phase is entered
in the Spherical model when the hypersphere radius exceeds $S_c$.

If this happens, the partition function of the constrained model
will be singular at criticality and, because of Eq. (\ref{gen.3}),
a point of non-analiticity will be found in the probability 
distribution $P(M,J)$ of the fluctuating variable ${\cal M}$. 

So far  we have discussed the case of equilibrium states, where
the r.h.s. of Eq. (\ref{gen.3}) can be interpreted as the partition
function of a restricted model. 
If the system is not in equilibrium, this expression is not 
amenable of the same interpretation. Nevertheless a singularity can still
be produced by a mechanism which resembles a dynamical phase-transition,
as it will be shown in Sec. \ref{VI}.

\subsection{Large volume limit} \label{III}

Introducing the integral representation of the $\delta$ function
$\delta(x) = \int_{\alpha-i\infty}^{\alpha + i\infty}\frac{dz}{2\pi i} \, e^{-zx}$,
Eq. (\ref{gen.3}) becomes
\be
P(M,J) =  \int_{\alpha - i\infty}^{\alpha + i\infty} \frac{dz}{2\pi i} \, e^{-zM} K_{\cal M}(z,J)
\label{gen.5}
\ee
where
\be
K_{\cal M}(z,J) = \int_{\Omega} d\varphi \, P(\varphi,J)e^{z{\cal M}(\varphi)}
\label{gen.6}
\ee
is the moment generating function of ${\cal M}$.
If the system is extended and ${\cal M}(\varphi)$ is an extensive macrovariable,
for large volume Eq.~(\ref{gen.5}) can be rewritten as
\be
P(M,J,V) =  \int_{\alpha - i\infty}^{\alpha + i\infty} \frac{dz}{2\pi i} \, e^{-V[zm + \lambda_{\cal M}(z,J)]}
\label{gen.66}
\ee
where we have explicitly separated the volume $V$ from the bunch of
control parameters $J$, $m$ is the density $M/V$ and
\be
-\lambda_{\cal M}(z,J) = \frac{1}{V}\ln K_{\cal M}(z,J,V)
\label{gen.7bis}
\ee 
is volume independent in the large-volume limit. 
Carrying out the integration by the saddle point method one arrives at
\be
P(M,J,V) \sim e^{-VI_{\cal M}(m,J)}
\label{gen.8}
\ee
with the rate function
\be
I_{\cal M}(m,J) = z^*m + \lambda_{\cal M}(z^*,J) 
\label{gen.9}
\ee
and where $z^*(m,J)$ is the solution, supposedly unique, 
of the saddle point equation
\be
\frac{\partial}{\partial z} \lambda_{\cal M}(z,J) = -m.
\label{gen.9bis}
\ee
Eq. (\ref{gen.8}) amounts to the large deviation principle,
according to which the probability of a fluctuation of a 
macrovariable is exponentially damped by the system volume 
with rate function $I_{\cal M}(m,J)$.  

\section{A specific example: The Gaussian model}
\label{IV}
 
As a simple, fully analytical model to test the above ideas
the Gaussian model was considered in~\cite{noi}. 
The set of microvariables are represented 
by a scalar order parameter
field $\varphi(\vec x)$, governed by the bilinear energy functional
\be
{\cal H}[\varphi] = \frac{1}{2} \int_V d \vec x \, [(\nabla \varphi)^2 + r \varphi^2 (\vec x)]
\label{GMD.1}
\ee
where $r$ is a non negative parameter.
In order to study both the equilibrium behavior and the non-equilibrium process
where time translational invariance is spoiled we consider a protocol
where the system is kept in equilibrium at the temperature $T_I$ at times
$t<0$. Then, at the time $t=0$ it is
instantaneously quenched to the lower temperature $T_F$. 
The dynamics, without conservation of the order parameter, is governed by the overdamped Langevin
equation~\cite{Goldenfeld}
\be
\dot{\varphi}(\vec x, t) = \left [ \nabla^2 - r \right ] \varphi(\vec x, t)  + \eta(\vec x, t)
\label{Gin.4}
\ee
where $\eta(\vec x, t)$ is the white Gaussian noise generated by the cold reservoir, 
with zero average and correlator 
\be
\langle \eta(\vec x,t)\eta(\vec {x}^{\prime},t^{\prime}) \rangle = 
2T_F \delta(\vec x -\vec {x}^{\prime})   \delta(t-t^{\prime}),
\label{LG.2}
\ee
where we have set to unity the Boltzmann constant.
Due to linearity, the problem can be diagonalized by Fourier transformation.
For the Fourier components
$\varphi_{\vec k} = \int_V d\vec x \, \varphi(\vec x) e^{i\vec k \cdot \vec x}$,
by imposing periodic boundary conditions,
one gets the equations of motion 
\be
\dot{\varphi}_{\vec k}(t) = -(k^2+r) \varphi_{\vec k}(t) + \eta_{\vec k}(t)
\label{FOUR.2}
\ee 
where the noise correlator is given by
\be
\langle \eta_{\vec k}(t) \eta_{\vec{k}^{\prime}}(t^{\prime}) \rangle  = 2T_F V \delta_{\vec k,-\vec{k}^{\prime}}\delta(t-t^{\prime}).
\label{FOUR.8}
\ee

\subsection{Fluctuations of a macrovariable}
\label{V}

In order to consider a specific example, let
us now focus on the following macrovariable 
${\cal M}(\{\varphi _k\}) = {\cal S}(\{\varphi _k\}) = 
\frac{1}{V}\sum_{\vec k}|\varphi_{\vec k}|^2$,
where $\{ \varphi _k\}$ indicates the whole set of
$\varphi _k$'s. 
Before proceeding let us stress that with this choice 
the restriction~(\ref{constraint})
amounts, in equilibrium, to the spherical constraint {\it \`a la}
Berlin-Kac, since it fixes the squared modulus of the order
parameter to a given value $S$. We expect, therefore, to
observe a singular point in the probability distribution 
$P(S)$. 
Out of equilibrium the restriction (\ref{constraint}) amounts to
force the order-parameter on the hypersphere at the time $t$ when
the observation is performed. 
This constrained model, therefore, is not related
to the properties of the dynamical spherical model~\cite{Godreche}, 
which requires the spherical
constraint to be imposed at all times and, 
to the best of our
knowledge, has never been considered before. 
Then one cannot, in principle, 
make any prediction based on the knowledge 
of the constrained model. 
We will see {\it a posteriori}, however,
that a singularity shows up also out of equilibrium.

According to the scheme of section~\ref{III},
all the information on the fluctuations of ${\cal S}(\{\varphi _k\})$ at the
generic time $t$ is contained in the rate function~(\ref{gen.9}), with $J=t$.
The evaluation of this quantity requires the preliminary computation of
the moment generating function. From the factorization property of the
Gaussian model and the
separability of ${\cal S}$ follows~\cite{noi} 
\be
K_{{\cal S}}(z,t) =   \prod_{\vec k}K_{{\cal S},\vec k}(z,t),
\label{EnHe.1}
\ee
with the single-mode factors given by
\be
K_{{\cal S},\vec k}(z,t) =  
\frac{1}{\sqrt{1 - \rho^{-1}_k(t)z}}
\label{EnHe.5}
\ee
where 
\be
\rho_k (t)= \frac{1}{2}\,\frac{k^2+r}{(T_I-T_F) e^{-2(k^2+r) t} + T_F}.
\label{ro.1}
\ee
The product on the r.h.s. of Eq. (\ref{EnHe.1}) is limited to wavevectors 
with $\vert \vec k \vert<\Lambda$, where $\Lambda $ is
an ultraviolet cutoff caused by the existence of a microscopic length scale
in the problem, like an underlying lattice spacing. 

Inserting Eq.~(\ref{ro.1}) into Eq.~(\ref{gen.7bis}), the saddle point equation~(\ref{gen.9bis})
can be written as
\be
s = \widetilde{F}_{{\cal S}}(z,t,V) 
\label{MDF.4}
\ee
where $s=S/V$ and the function in the right hand side is given by
\be
\widetilde{F}_{{\cal S}}(z,t,V) = \frac{1}{V} \sum_{\vec k} 
\frac{1}{2[\rho_k(t) - z]}.
\label{MDF.6}
\ee

Transforming the sum in Eq.~(\ref{MDF.6}) into an integral, the saddle point equation~(\ref{MDF.4}) 
can be rewritten as
\be 
s=F_{\cal S}(z,t)
\label{F.2}
\ee 
with
\be
F_{\cal S}(z,t) = \frac {\Upsilon_d}{2} \int_0^{\Lambda} \frac{dk}{(2 \pi)^d} \, \frac{k^{d-1}}{\rho_k(t) - z}
\label{F.1}
\ee
where $d$ is the space dimensionality, $\Upsilon_d=2\pi^{d/2}/\Gamma (d/2)$ is the $d$-dimensional solid angle and
$\Gamma$ the Euler gamma function. 
The formal solution is given by
\be
z^*(s,t) = F_{{\cal S}}^{-1}(s,t)
\label{sol.1}
\ee
where $F_{{\cal S}}^{-1}$ 
is the inverse, with respect to $z$, of the function defined by Eq.~(\ref{F.1}).
The existence of this solution
depends on the domain of definition of  $F_{\cal S}^{-1}$.
Since ${\cal S}$ is positive, and from Eqs. (\ref{ro.1}) it is
easily verified that the minimum of $\rho _k(t)$ is at $k=0$ at any time,
$F_{\cal S}^{-1}$ is defined for $z \leq \rho_0(t)$. Then
\be
F_{\cal S}(z,t) \leq s_C(t)
\label{sol.2}
\ee
with
\be
s_C(t) = F_{\cal S}(z=\rho_0,t). 
\label{uppb.1}
\ee
According to Eq. (\ref{ro.1}), 
\be
[\rho_k(t) - \rho_0(t)]\simeq C_tk^2 \hspace{1cm},\hspace{1cm}k^2t\ll 1,
\label{vanish}
\ee 
vanishes with $k$ like $k^2$ (with $2 C_t= [T_F + (T_I - T_F) e^{-2rt}(1+ 2 r t)]/[T_F + (T_I - T_F)e^{-2rt}]^2$).
Then for $d>2$ 
the singularity is integrable on the r.h.s. of Eq. (\ref{F.1}), $s_C(t)$ is finite and the solution~(\ref{sol.1}) 
exists only for $s \leq s_C(t)$. 
In order to find the solution for 
$s > s_C(t)$ one must proceed following an analytical treatment 
similar to that of 
Berlin and Kac~\cite{BK}. This will be done in Sec. \ref{stde}.

\begin{figure}[h]

	\vspace{1cm}

    \centering
   \rotatebox{0}{\resizebox{.4\textwidth}{!}{\includegraphics{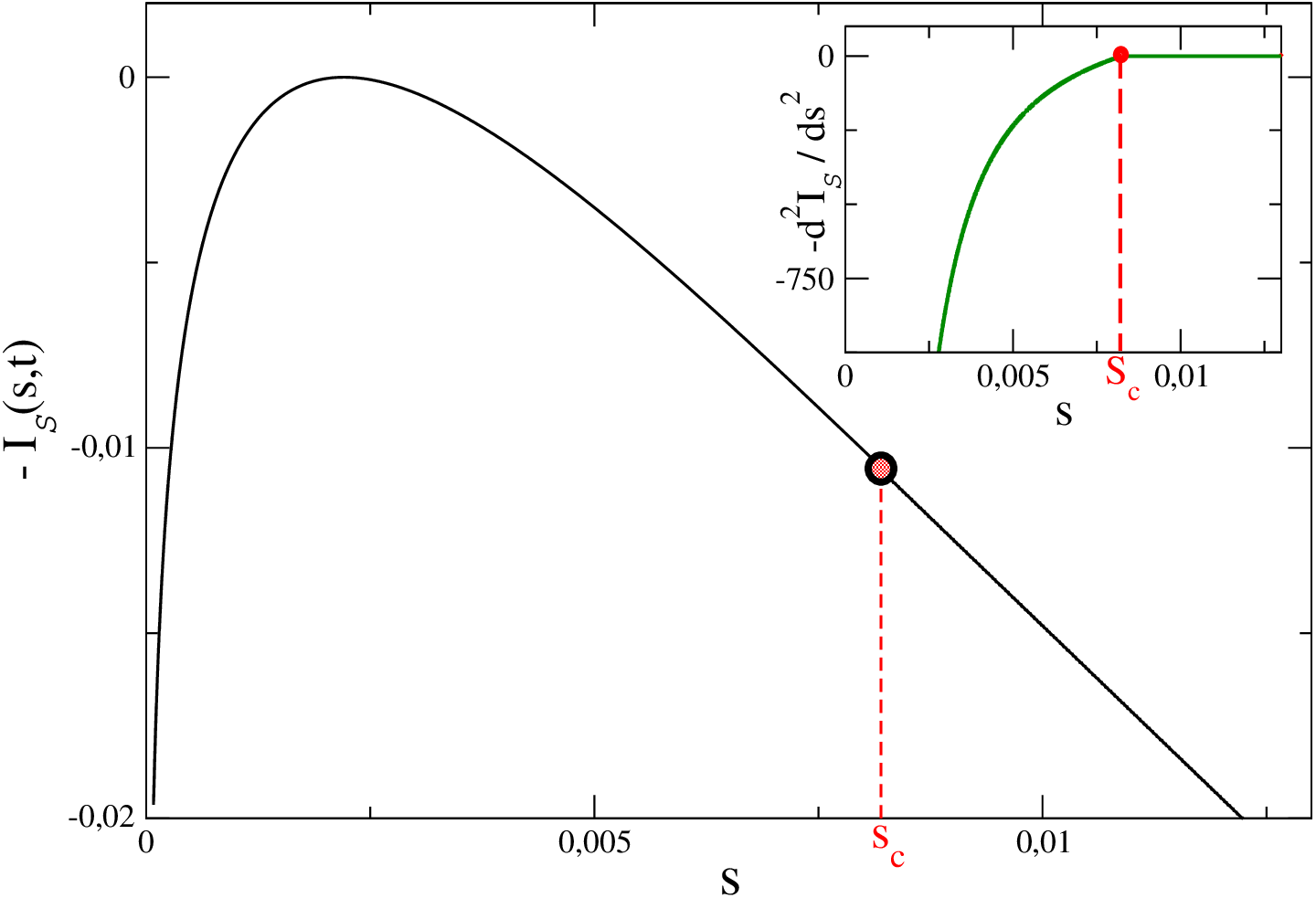}}}\vskip 0.7cm

   \caption{Rate function $-I_{\cal S}(s,t)$ for the sample variance. 
Parameters $\mu_k=2, t=2, r=1,T_I=1,T_F=0.2$, $d=3$.
In the inset the second derivative is shown.}
\label{ratef}
\end{figure}

Alternatively, as done in \cite{noi}, one can proceed in a more physically 
oriented way, as 
usually done in the standard treatment of BEC~\cite{Huang}. 
This amounts to separate the $k=0$
term from the sum and rewriting Eq.~(\ref{F.2}) as 
\be
s = \frac{1}{V} \cdot \frac{1}{2[\rho _0(t)-z^*]} + F_{\cal S}(z^*,t).
\label{4}
\ee
Then, $s_C(t)$ defines a critical line on the
$(t,s)$ plane separating the normal phase (below) from the condensed phase (above).
Below, the first term in the right hand side of
Eq.~(\ref{4}) is ${\cal O}(1/V)$ and negligible, while above it takes
the finite value $[s - s_C(t)]$, due to the {\it sticking}~\cite{Huang,BK}
of $z^*$ to the $s$-independent value $z^*=\rho_0(t)$. Summarising,
\be
z^*(s,t)  = \left \{ \begin{array}{ll}
        F_{\cal S}^{-1}(s,t) ,\;\; $for$ \;\; s \leq s_C(t),\\
        \rho_0(t) ,\;\; $for$ \;\;  s > s_C(t).
        \end{array}
        \right .
        \label{ODG.6}
        \ee
Plugging this determination of $z^*$ into Eq.~(\ref{gen.9}), and
Eqs.~(\ref{EnHe.1},\ref{EnHe.5}) in Eq.~(\ref{gen.7bis}) one 
arrives to an explicit determination of the rate function $I_{\cal S}(s,t)$.
This quantity is plotted in Fig. \ref{ratef}. 

As it suggested by the twofold expression~(\ref{ODG.6}) of the saddle point
solution, and how it will be explicitly shown in Sec.~\ref{sing},
$I_{\cal S}$ is singular at $s=s_C(t)$.
The treatment above clearly shows that the mechanism whereby the singularity
is built is a condensation transition, in analogy with the case of the BEC.
As already discussed, the difference with BEC is that here condensation is
not observed as a typical event but in the fluctuation of a macrovariable.
Let us add that $s_C$ corresponds to a very rare fluctuation, 
since as we will show in the next section $s_C(t)$ 
is much larger than the average value.

\subsection{Steepest descent} \label{stde}

In this section we discuss in detail the steepest descent evaluation of the 
integral in Eq.~(\ref{gen.66}).

For $s\le s_C$ the saddle point equation~(\ref{F.2}) admits a solution 
$z=z^*(s)$.

In the region $s>s_C$ 
the evaluation of the integral 
for the probability~(\ref{gen.66}) can be
done using analytical tools inspired to those developed in~\cite{BK}.
Let us consider this integral (with $M=S$ and $J=t$) 
in the neighbourhood  of the origin of the cut  extending in the 
$z$-plane from $z=\rho_0$ to $\infty$. Singling out the contribution 
of the  $k=0$ mode Eq.~(\ref{gen.66}) reads
	\be
	P(S,t,V) = \int_{-i\infty}^{i\infty}\frac{dz}{2\pi
 i}e^{-V[f(z,s,t)]}\sqrt{\frac{\rho_{0}(t)}{\rho_{0}(t)-z}}
	\label{stter}
	\ee
	 with
	\be
	f(z,s,t)= sz + \lambda^{0}_{\cal S}(z,t),
        \label{stq}
        \ee
\be
\lambda^{0}_{\cal S}(z,t) = 
	\frac{1}{2}\int \frac{d^d k}{(2 \pi)^d} \ln[ {1 - \rho^{-1}_k(t)z}], 
\label{stqq}	\ee 
	and
\be
 \frac {\partial \lambda^0_{\cal S}(z,t)} {\partial z} = - {F}_{{\cal S}}(z,t). 
\label{st1bis}
\ee
Since $\lambda^0_{\cal S}(z,t)$ is analytical in the cut plane, its behaviour in the neighborhood   
  of $\rho_0(t)$ can be obtained by  analytic continuation of ${F}_{{\cal S}}(z,t) $ and then  integration of (\ref{st1bis}).  

By using the representation
\be
\frac{1}{\rho_k - z}=  \int_0^\infty e^{x(z-\rho_k)}dx=-\int_0^\infty
e^{x(z-\rho_0)+x(\rho_0-\rho_k)}dx,
\label{st2}
\ee
where here and in the following for simplicity we will drop the
time dependence in $\rho _k(t)$,
one can write
\be
2F_{\cal S}(z,t) = \frac{\Upsilon _d}{(2\pi)^d} \int_0^\infty
dk k^{d-1} \int_0^\infty e^{x\Delta _z+x(\rho_0-\rho_k)}dx
=\frac{\Upsilon _d}{(2\pi)^d}\int_0^\infty dx e^{x\Delta _z}\int_0^\infty
k^{d-1}e^{x(\rho_0-\rho_k)}dk,
\label{st22}
\ee
where $\Delta _z=z-\rho_0$.  Then one has
\be
2(2\pi )^dF_{\cal S}(z,t)/\Upsilon _d\equiv  \int_0^\infty e^{x\Delta _z}I(x)dx
\label{st3}
\ee
with 
\be
I(x) = \int_0^\infty k^{d-1}e^{x(\rho_0-\rho_k)}dk.
\label{st4}
\ee
For $x\gg\frac{2t}{C_t}$, where $C_t$ is defined in Eq.~(\ref{vanish}),
one has 
\be 
I(x)\simeq
\frac{x^{-d/2}\Gamma(d/2)(C_t)^{-d/2}}{2}.
\label{st5}
\ee
We now evaluate  the derivative $W'(\Delta _z)$  
of the quantity
$W(\Delta _z)=\int_0^\infty e^{x\Delta _z}I(x)dx$ on the r.h.s. of Eq.~(\ref{st3}).
For small $\Delta _z$,
from Eq.~(\ref{st5}) one has
 $W'(\Delta _z) = A(-\Delta _z)^{d/2-2}$ where  $A=\int_0^\infty
y^{-d/2+1}e^{-y}\frac{\Gamma(d/2)}{2(C_t)^{d/2}}dy$ for  $2<d<4$. 
Then   $W(\Delta _z)$ can be calculated by  integration and Eq.(\ref{st3}) becomes
\be
F_{\cal S}(z,t)  = s_C -   \frac {\Upsilon _d A}
{(2\pi)^d (d-2)}(-\Delta _z)^{d/2-1}, 
\label{st6}
\ee
where $s_c$ is a shorthand for $s_C(t)$.
Therefore in the neighborhood of
$\rho_0$ the derivative of $f(z,s,t)$ can be written
as
\be
f'(z,s,t)=s-s_C+\frac{\Upsilon _d A}{(2\pi)^d(d-2)}(\rho_0-z)^{d/2 -1}
\label{st7}
\ee
and hence
\be
f(z,s,t)=f(\rho_0,s,t)+(s-s_C)(z-\rho_0)+K(\rho_0-z)^{d/2} 
\label{st8}
\ee
where $K=-\frac{2\Upsilon _d A}{(2\pi )^dd(d-2)}<0$. The above expansion shows 
that the integrand in Eq.(\ref{stter}) always has a saddle point at $z=\rho_0$ if $s > s_C$, namely the statement of Eq.~(\ref{ODG.6}),
with the steepest descent contour having a cusp in $z=\rho_0$.
Hence the dominant contribution to the integral in Eq.~(\ref{stter}) can be now evaluated as
\begin{eqnarray}
P(S,t,V)&=& \frac{e^{-Vf(\rho_0,s,t)}\sqrt{\rho_{0}}}{2\pi i}
\int_{-i\infty}^{i\infty}\frac{e^{V(s-s_C)(\rho_0-z)}}{(\rho_0-z)^{1/2}}dz
=
e^{-Vf(\rho_0,s,t)}
\sqrt{\frac{\rho_{0}}{\pi V(s-s_C)}} \nonumber \\
&=& 
\sqrt{\frac{\rho_{0}}{\pi V(s-s_C)}} \,
e^{-V\rho_0[\lambda^0_{\cal S}(\rho_{0},t)+s]}
,
\label{st9}
\end{eqnarray}
where the integral representation of the $\Gamma$-function has been used.
This shows that the rate function is linear in $s$, as it is
clear from Eq.~(\ref{gen.9}) (with $m=s$) when $z^*$ is independent
on $s$. 

\subsection{Singularity} \label{sing}

The previous calculation shows that
there is singularity (the marked dot in Fig. \ref{ratef}) located
at $s=s_C$ in the rate function.

In order to analyze the nature of such singularity 
we will compute $s$-derivatives of the rate 
function on the right and on the left of $s_C$, and then take the 
limit for $s\to s_C^\pm$.
Starting with the sector $s>s_C$, 
due to the {\it sticking} $z^*=\rho_0$ of the saddle point solution, 
from Eq.~(\ref{stq}) one has
\be
{\frac{df(z^*(s),s,t)}{ds}}=\rho_0
\label{dis1}
\ee
while all  right derivatives of higher order are zero. 

On the other hand, for $s < s_C$,
using Eqs.~(\ref{stq},\ref{st1bis}) one has
\be
{\frac{df(z^*(s),s,t)}{ds}}=
\left [z+z'(s-F_{\cal S}(z,t))\right ] _{z=z^*(s)}=
z\vert_{z=z^*(s)}=z^*(s),
\label{dis2}
\ee
where $z'=dz(s)/ds$ and we have used the saddle point equation~(\ref{F.2}). For $s\to s_c$, $z\to \rho_0$ and this left
derivative equals the right one (\ref{dis1}).
The second left derivative is given by
\be
\left . {\frac{d^2f(z^*(s),s,t)}{ds^2}}\right \vert_{s_C^-}  =
\left . z'\right \vert_{z=z^*(s)}.
\label{dis3}
\ee
Near $s_C$, for $z\to \rho_0$, using Eqs. (\ref{F.2},\ref{st6})
one has that $z'\vert _{z=z^*(s)}$ goes to zero as 
$(\rho_0-z)^{2-d/2}$ for $d<4$. Hence the second left derivative
vanishes and equals the right one at $s=s_C$.
This is shown in the inset of Fig.~\ref{ratef}.
This figure shows that the second derivative has a kink at $s=s_C$,
as it can be shown by considering the 
third left derivative, which reads
\be
{\frac{d^3f(z^*(s),s,t)}{ds^3}}
=z''\vert _{z=z^*(s)}.
\label{dis6}
\ee
Using again Eqs. (\ref{F.2},\ref{st6}) for $s$ near $s_C$ one has
$\frac{d^3f(z^*(s),s,t)}{ds^3}=\frac{-2(2\pi)^d (d-4)}
{\Upsilon _d A^2}(\rho_0-z)^{3-d}$.
Therefore in $d=3$ the large deviation function has a discontinuity 
on the third derivative at $s=s_c$.

\subsection{Phase diagram}
\label{VI}

\begin{figure}[h]

	\vspace{1cm}

    \centering
   \rotatebox{0}{\resizebox{.35\textwidth}{!}{\includegraphics{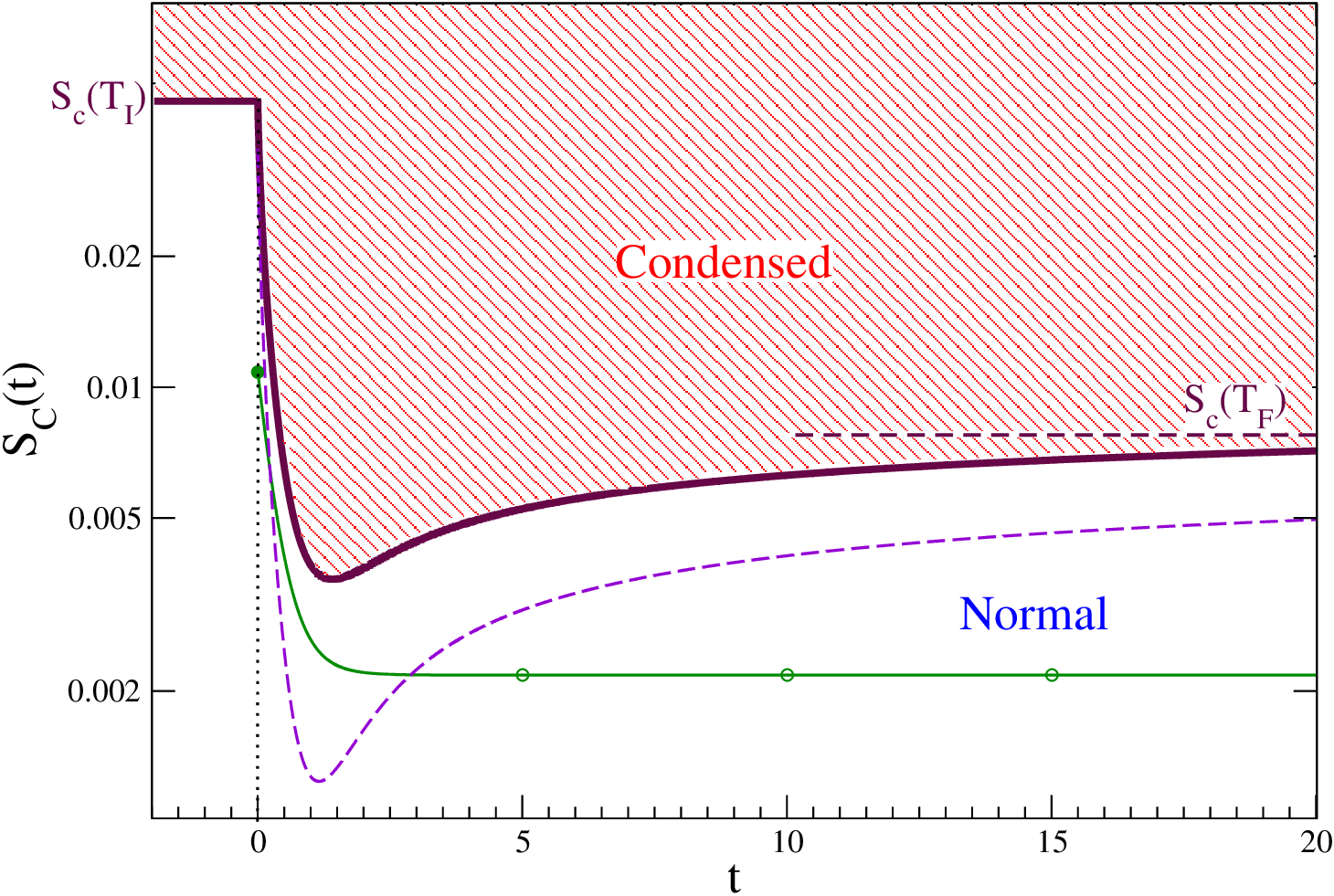}}}\vskip 0.7cm
   \rotatebox{0}{\resizebox{.35\textwidth}{!}{\includegraphics{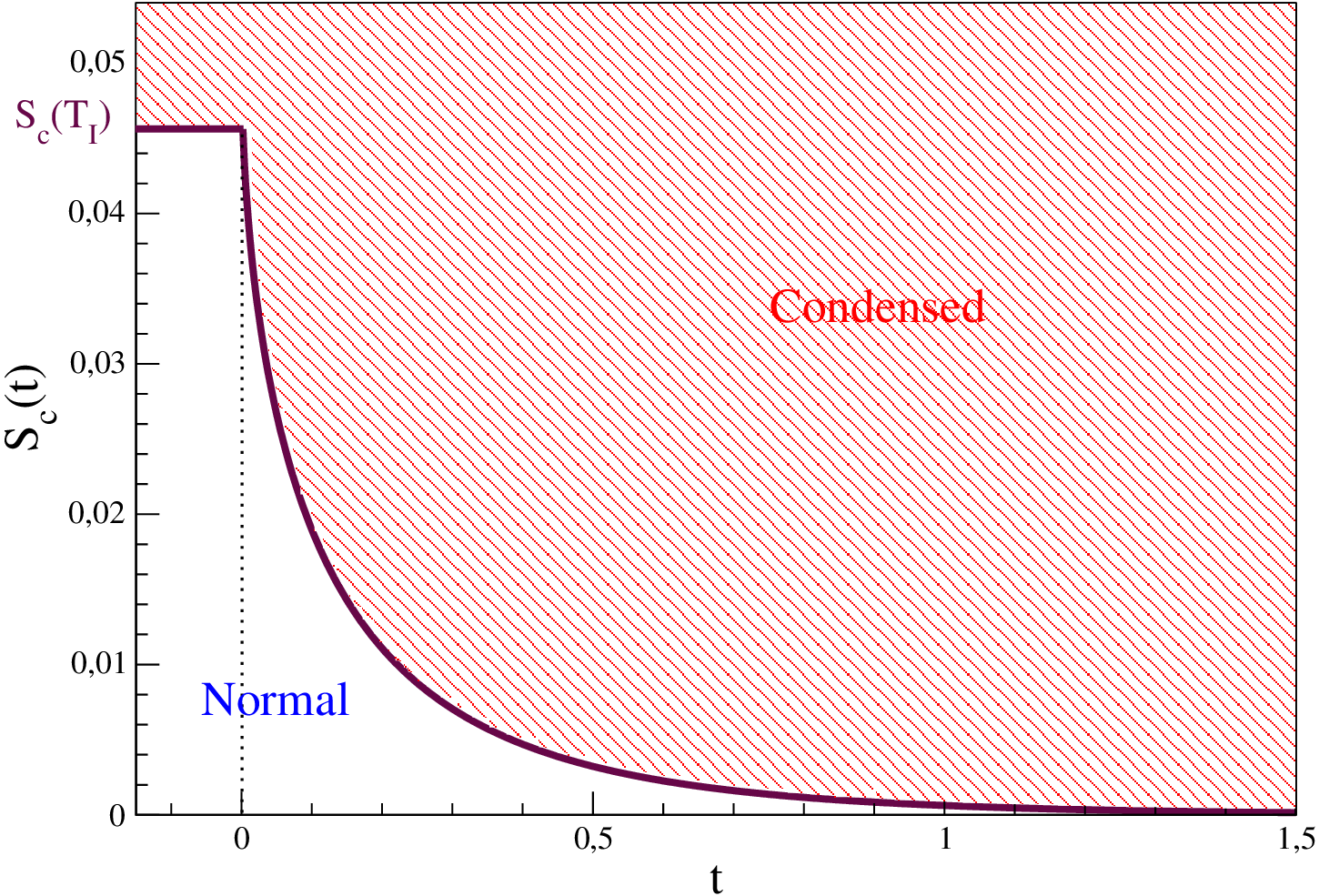}}}\vskip 0.7cm

   \caption{Phase diagram of order parameter sample variance.
The upper horizontal dashed line corresponds to $s_C(T_F)$. The green line is the plot of
$\langle s(t) \rangle$. The parameters are: $r=1,T_I=1,d=3$ and $T_F=0.2$ 
or $T_F=0$ in the upper and lower panel respectively. In the upper panel the 
lower dashed line is the difference $[s_C(t) - \langle s(t) \rangle]$.}
\label{ph_diag_S}
\end{figure}

The critical line $s_c(t)$ for $d=3$ 
is displayed in Fig.~\ref{ph_diag_S}.
In order to understand this phase diagram, 
one should keep in mind that fixing the value of $s$ amounts to implement
a spherical constraint {\it \`a la} Berlin and Kac~\cite{BK}. 
Let us first consider equilibrium, in the time region $t \leq 0$ preceding
the quench. Here, the critical line 
is horizontal and corresponds to the critical threshold $s_C(T_I)$ of the spherical
model at the temperature $T_I$.

Consider, next, the relaxation regime after the quench, for $t > 0$.
Following the discussion in~\cite{noi},
when quenches to a finite
final temperature are considered (upper panel of Fig.~\ref{ph_diag_S}) 
there are two time regimes separated by the minimum of the critical line,
about the characteristic time $\tau \sim r^{-1}$, which is 
the relaxation time of the slowest mode with $\vec k=0$. 
In the first regime $(0 < t < \tau)$ the system
is strongly off equilibrium and the threshold $s_C(t)$ drops abruptly.
In the second regime $(t > \tau)$ the system gradually equilibrates
to the final temperature and $s_C(t)$ saturates slowly toward the final equilibrium 
value $s_C(T_F) < s_C(T_I)$.
A few observations are in order: 
i) The plot of the average $\langle s(t) \rangle$ lies
below the critical line, showing that condensation
of fluctuations is always a rare event. 
The plot of $[s_C(t) - \langle s(t) \rangle]$ shows
that the {\it rarity} of the condensation event varies with time and that the most favourable
time window for condensation is around $\tau$, where the difference is minimized.
Hence, condensation of the fluctuations is enhanced by the off equilibrium dynamics.
ii) The non-monotonicity of the critical line is a remarkable dynamical feature, leading
to a re-entrance phenomenon. Namely, when the transition is driven by $t$, 
and $s$ is kept fixed to a value in between $s_C(T_F)$ and $s_C(T_I)$, a fluctuation of this size at first
is normal and  then condenses,  while for $s$ in between the minimum of the critical line and $s_C(T_F)$, 
the fluctuation undergoes a second and reverse transition becoming  normal again at late times.

For a quench to $T_F=0$, shown in the lower panel of Fig. \ref{ph_diag_S},
the critical value in the equilibrium state at
the final temperature $s_C(T_F=0)=0$ vanishes and condensation is asymptotically
observed for any allowed value of $s$.

\section{Conclusions}
\label{VII}

In this paper we have discussed a mechanism whereby a singularity can
be produced in the probability distribution $P(M)$ of the fluctuations of a
macrovariable ${\cal M}$ in a generic thermodynamic system.
We have addressed this problem in the framework of a specific model,
the Gaussian model, where analytical calculations can be carried over and
the non-analytical behavior can be explicitly exhibited. 
We have discussed how this phenomenon can be related to the behavior 
of a dual model obtained from the original one
by exerting a constraint. For the Gaussian model considered here, 
upon choosing
${\cal M}$ as the order-parameter variance ${\cal S}=\varphi ^2(\vec x,t)$, 
the dual model is represented by the Spherical model. Then, at least in 
equilibrium, the singularity of $P(S)$ can be interpreted as the
counterpart of the ferro-paramagnetic phase-transition determined
by the spherical constraint. An analogous phenomenon, with a similar
interpretation, is observed also out of equilibrium, as we have shown 
by considering the evolution of the Gaussian model after a temperature quench.
The non-analytical behavior observed in this model has been
related to the properties of the steepest descent path whose evaluation
was carried over explicitly in this paper.

The phenomenon of condensation of fluctuations
is not restricted to the Gaussian model neither to the kind of
quantities we have considered insofar but is a much more
general property. It has been observed, for instance, considering the
fluctuations of the heat exchanged by a mean-field model 
(an attempt to go beyond mean-field was done in~\cite{CGP}) 
of a ferromagnet in non-equilibrium conditions 
and the thermal bath in~\cite{US}.  
Moreover, in~\cite{FL} the singular behavior of the fluctuations of
composite operators whose average yield the correlation and the response
function was shown.
Finally, although for computational purposes we have considered the case
of a non-interacting system, the mechanism whereby the fluctuation spectrum
can be related to the properties of a dual system  is a general property. 
An interesting future work, therefore, should be devoted
to the investigation of fluctuations in interacting systems.

\noindent {\it e-mail addresses} - corberi@sa.infn.it, gonnella@ba.infn.it, AntPs@ntu.edu.sg


\begin{thebibliography}{99}
 
\bibitem{Huang}
K. Huang, {\it Statistical Mechanics}, John Wiley and Sons, New York 1967

\bibitem{BK}
T. H. Berlin and M. Kac, Phys. Rev. {\bf 86}, 821 (1952)

\bibitem{LN}
For the condensation transition when the spherical constraint is imposed in the mean
via the large $N$ limit, see
C. Castellano, F. Corberi, and M. Zannetti, Phys. Rev. E {\bf 56}, 4973 (1997)

\bibitem{Goldenfeld}
N. Goldenfeld, {\it Lectures on Phase Transitions and the Renormalization Group}, Addison-Wesley Publishing Co.,
Reading, Mass. 1992;
P. M. Chaikin and T. C. Lubenski, {\it Principles of Condensed Matter Theory}, Cambridge University Press 1995;
P. C. Hoenberg and B. I. Halperin, Rev. Mod. Phys. {\bf 49}, 435 (1977).

\bibitem{Touchette}
H. Touchette, Phys. Rep. {\bf 478}, 1 (2009).

\bibitem{schutz}
R.J. Harris, A. R\'akos, and G.M. Schuetz, J. Stat. Mech. P08003 (2005)

\bibitem{Kafri}
N. Merhav and Y. Kafri, J. Stat. Mech. P02011 (2010)

\bibitem{FL}
F.Corberi and L.F.Cugliandolo, J. Stat. Mech. P11019 (2012).

\bibitem{US}
F. Corberi, G. Gonnella, A. Piscitelli and M. Zannetti, J. Phys. A: Math. Theor. {\bf 46}, 042001 (2013)

\bibitem{Gambassi}
A. Gambassi and A. Silva, Phys. Rev. Lett. {\bf 109}, 250602 (2012).

\bibitem{Evans}
J. Szavits-Nossan, M. R. Evans and S. N. Majumdar, Phys. Rev. Lett. {\bf 112}, 020602 (2014).






\bibitem{noi}
M. Zannetti, F. Corberi, G. Gonnella  arXiv:1404.3975. 
M. Zannetti, F. Corberi, G. Gonnella and A. Piscitelli, submitted to
Communications in Theoretical Physics. 

\bibitem{ritort}
A. Crisanti and F. Ritort, Europh. Lett. {\bf  66}, 253 (2004).

\bibitem{Ciliberto}
J. R. Gomez-Solano, A. Petrosyan and S. Ciliberto, Phys. Rev. Lett. {\bf 106}, 200602 (2011)

\bibitem{Godreche}
C. Godr\`eche and J. M. Luck, J. Phys. A: Math. Gen. {\bf 33}, 9141 (2000)
F. Corberi, E. Lippiello and M. Zannetti, Phys. Rev. E, {\bf 65} 046136 (2002);
A. Annibale and P. Sollich, J. Phys. A: Math. Gen. {\bf 39}, 2853 (2006).

\bibitem{derrida}
F. Ritort, J. Stat. Mech.: Theory and Experiment, P10016 (2004);
B. Derrida, J. Stat. Mech. P07023 (2007); C. Jardina, J. Kurchan and L. Peliti, Phys. Rev. Lett. {\bf 96},
120603 (2006); C. Jardina, J. Kurchan, V. Lecomte and J. Tailleur, J. Stat. Phys. {\bf 145}, 787 (2011).






\bibitem{Amit}
D. J. Amit and M. Zannetti, J. Stat. Phys. {\bf 7}, 31 (1973).

\bibitem{Zin}
J. Zinn-Justin, {\it Quantum Field Theory and Critical Phenomena}, Chpt. 30, 4th Edition, Clarendon Press, Oxford
(2002).

\bibitem{Ma}
S. K. Ma, {\it Modern Theory of Critical Phenomena}, Chpt. IX, W. A. Benjamin Inc., Reading, Mass. (1976);

\bibitem{heat}
This type of transition was first observed in the fluctuations of the heat exchanged
by a ferromagnet quenched below the critical point in Ref.~\cite{US} and in the 
fluctuations of composite operators whose average are correlation and response
functions in Ref.~\cite{FL}. 

\bibitem{CGP}
F. Corberi, G. Gonnella, A. Piscitelli
J. Stat. Mech. (2011) P10022.


\end{thebibliography}
\end{document}